\documentclass[
reprint,
amsmath,
amssymb,
aps,
prl
]{revtex4-2}

\usepackage{graphicx}
\usepackage{dcolumn}
\usepackage{bm}

\usepackage{color}
\usepackage{braket}

\begin{document}

\title{Quantum Particle Statistics in Classical Shallow Water Waves}%

\author{Idan Ceausu}
\author{Yuval Dagan}%
 \email{yuvalda@technion.ac.il\\}
\affiliation{%
 Faculty of Aerospace Engineering, Technion - Israel Institute of Technology, Haifa, 320003, Israel.\\
}%


\begin{abstract}
We present a new hydrodynamic analogy of nonrelativistic quantum particles in potential wells. Similarities between a real variant of the Schr\"odinger equation and gravity-capillary shallow water waves are reported and analyzed. We show that when locally oscillating particles are guided by real wave gradients, particles may exhibit
trajectories of alternating periodic or chaotic dynamics while increasing the wave potential. The particle probability distribution function of this analogy reveals the quantum statistics of the standard solutions of the Schr\"odinger equation and thus manifests as a classical deterministic interpretation of Born's rule. Finally, a classical mechanism for the transition between quasi-stationary states is proposed.
\end{abstract}

\maketitle

Since the introduction of the Madelung transformation~\cite{Madelung1926} and subsequent interpretations by Broglie and Bohm~\cite{Bohm1952}, numerous studies have revealed a close relation between hydrodynamics and quantum mechanics.
Recent advances in hydrodynamic quantum analogies revealed the emergence of quantum-like statistics for particles interacting in a deterministic hydrodynamic framework.
One of the most successful was found by Couder and Fort, who experimentally observed millimetric oil droplets bouncing over a vibrating bath that feature the statistical behavior of many quantum mechanical systems~\cite{Couder2006, bush2020hydrodynamic}. 
In this hydrodynamic quantum analogy (HQA), droplets interact in resonance with a quasi-monochromatic wavefield they generate and exhibit a self-propelling mechanism. This analog has extended the range of classical physics to include many features previously thought to be exclusively quantum, including tunneling \cite{eddi2009unpredictable}, Landau levels~\cite{fort2010path,oza2014pilot}, the quantum harmonic oscillator~\cite{perrard2014self}, the quantum corral~\cite{gilet2016quantumlike,cristea2018walking}, the quantum mirage~\cite{saenz2018statistical}, and Friedel oscillations~\cite{saenz2020hydrodynamic}.
However, the waveform in this analogy is not identical to the quantum mechanical wave function, which, for example, exhibits different phase and group speed relations. Furthermore, the interpretation and the appearance of quantum-like statistics of these analogies are usually limited to a strobed particle-wave framework~\cite{bush2020hydrodynamic}.

Therefore, it is instructive to seek waveforms that conform to both hydrodynamics and quantum mechanics.
Notable studies deal with classical mechanics to interpret both relativistic and nonrelativistic quantum mechanical wave equations~\cite{shinbrot2019dynamic, Borghesi2017}.
The Klein-Gordon (KG) equation may serve both as a real wave field in relativistic quantum field theory and as a linear model for long gravity water waves, as well as in recent hydrodynamic analogies~\cite{Drezet2020}. 
Recently, a hydrodynamically-inspired quantum dynamics theory was developed by~\citet{dagan2020hydrodynamic}, a conceptual model of relativistic quantum dynamics inspired by de Broglie's pilot wave theory.
In this framework, the particle is assumed to be a localized—yet infinite—oscillating disturbance, externally forcing a KG wave equation. A relativistic dynamic equation couples the motion of the localized particle to the wave.
Using this deterministic framework, several features of quantum mechanics were revealed. The most intriguing is probably the particle momentum in this analogy, which is associated with inline oscillations corresponding to the relation $p=\hbar k$, realized through interactions with the wave field. Notably, the particle speed modulations are averaged at the de Broglie wavelength and modulated by the relativistic frequency $k c$.
Similar observations were reported by \citet{durey2020hydrodynamic}, who revealed the wave generation and self-propelling mechanism for the coupled wave-particle system and provided a fundamental analytical validity to the subsequent work on the hydrodynamic field theory. More recently, inline particle oscillations comparable to de Broglie’s wavelength formula were realized through exact solutions of a fully classical dynamic model \cite{dagan2023relativistic, 2024pilotwave}. However, mainly due to large-scale separation and extremely high frequency associated with the Compton scale, nonrelativistic dynamics were not resolved using this approach. 

In this letter, we propose a hydrodynamic analogy based on a real variant of the Schr\"odinger equation, where the particle is guided by wave gradients, thus offering a deterministic interpretation for nonrelativistic quantum particle dynamics. 

In its most general form, the wave function $\psi$ is a complex function, which, according to the Born rule, manifests as the probability density of the particle's location for the real-valued function $|\psi|^2$. Here, on the other hand, we formulate a real wave field by splitting the Schr\"{o}dinger equation into two coupled equations using the complex notation $\psi=\xi+i\theta$,
\begin{equation} \label{coupled_u_v}
\left\{ \begin{aligned} 
  \frac{\partial \xi}{\partial t} &= -L\theta ~;\\
  \frac{\partial \theta}{\partial t} &= L\xi~,
\end{aligned} \right.
\end{equation}
where $L=\frac{1}{\hbar}\left(\frac{\hbar^2}{2m}\frac{\partial^2}{\partial x^2}-V\right)$ is an operator satisfying the complex Schr\"{o}dinger equation $\left(\frac{\partial}{\partial t}-iL\right)\psi=0$.
Assuming that the potential $V$ is time-independent, one can differentiate in time the first equation of \eqref{coupled_u_v} and substitute it in the second to get a wave equation for the real part of $\psi$,
\begin{equation} \label{u_wave_equation}
    \frac{\partial^2 \xi}{\partial t^2}=-L^2 \xi~.
\end{equation}
Notably, Schr\"{o}dinger himself referred to this equation and presented it as the ``real wave equation", initially preferring it over the well-known complex equation~\cite{schrodinger2003collected}. However, he was not able to find a real equation for a non-conservative system, where the potential $V$ is time-dependent; he wrote:``Meantime, there is no doubt a certain crudeness in the use of a complex wave function ... I believe, of the very much more congenial interpretation that the state of the system is given by a real function and its time-derivative ... a real wave equation of probably the fourth order, which, however, I have not succeeded in forming for the non-conservative case." \cite{schrodinger2003collected}.

\par We further investigate the real variant of the Schr\"{o}dinger equation and simplify our analysis by assuming spatial piecewise constant potentials, $V=V_0(x)$, as in the cases of a particle in a box, and under the effect of potential wells. By expanding the RHS expression of equation \eqref{u_wave_equation}, we may write a fourth-order wave equation,
\begin{equation} \label{the wave u}
    \xi_{tt}=-\frac{\hbar^2}{4m^2}\xi_{xxxx}+\frac{V_0}{m}\xi_{xx}-\frac{V_0^2}{\hbar^2}\xi~.
\end{equation}
We will refer to this equation as the \textit{real form} of the conservative constant potential Schr\"{o}dinger equation (RS). A new dispersion relation may then be extracted,
\begin{equation} \label{dr QM}
    \omega^2=\frac{\hbar^2}{4m^2}k^4+\frac{V_0}{m}k^2+\frac{V_0^2}{\hbar^2}~.
\end{equation}
Note that this dispersion relation can also be derived simply by squaring the classical energy of a particle
and implement de broglie’s hypothesis, $(E,p)=\hbar\cdot(\omega,k)$. 
And so, we may infer that any solution to the Schr\"{o}dinger equation is a solution to the higher order RS equation. In our model, we compare the RS equation to surface waves. The RS equation can describe, as we shall see, the elevation of the free fluid interface in shallow water waves in its hydrodynamic form.

From the dispersion relation of the RS equation, we observe that the kinetic energy of the system does not solely depend on the difference between the energy and the constant potential, $E-V_0$.
In the absence of a potential, however, the dispersion relation of the RS equation is precisely that of the original complex one. 

We shall now present the free surface wave dispersion relations and the shallow water wave equation of our hydrodynamic analogy.
Consider free surface waves denoted by $z = \eta(x,y,t)$, where $\vec{x}=(x,y)$ is the horizontal plane and $z = -H$ is the bottom floor boundary. 
Assuming incompressible, inviscid, potential flow, $\vec{u} = \nabla \phi$, and small surface slopes, we can linearize the Euler equations with the stress balance condition at the free surface and expand $\eta(x,y,t)$ and $\phi(x,y,z,t)$ using Fourier series,
$\eta = \sum_{m=0}^{\infty} a_m(t)e^{i\vec{k}_m\cdot\vec{x}}$ and $\phi = \sum_{m=0}^{\infty} f_m(z)b_m(t)e^{i\vec{k}_m\cdot\vec{x}}$, where $\vec{k}_m=(k_{x_m},k_{y_m})$. By substituting into the linearized equations the shape function $f_m(z) = \frac{\cosh{(k_m(z+H))}}{\sinh{(k_mH)}}$, we may write the gravity-capillary (GC) dispersion relation,
\begin{equation} \label{GC dispersion}
    \omega_F^2 = \tanh{(k_FH)}\left(\frac{\sigma}{\rho}k_F^3 + gk_F \right),
\end{equation}
where $\omega_F,k_F$ will be referred to as the Faraday natural frequencies.
In the limit of shallow liquid, $k_mH \ll 1$, and $\tanh{(k_F H)} = k_F H + \mathcal{O}\left( (k_F H)^3 \right)$, the shallow-water GC dispersion
\begin{equation} \label{shallow GC dispersion}
    \omega_F^2 = \frac{\sigma H}{\rho}k_F^4 + gHk_F^2~,
\end{equation}
is derived.
Under the shallow water limit assumption, we may also write an explicit expression for the free surface wave equation; further assuming a two-dimensional configuration and a one-dimensional interface, $z=\eta(x,t)$, we have $f(z) \approx \frac{1}{k_m H}$. 
$\phi$ is now independent of $z$. Integrating the continuity equation over the depth to get a fourth-order wave equation for the free surface of the bath, 
we can write an explicit wave equation for shallow water waves, 
\begin{equation}
    \eta_{tt} = -\frac{\sigma H}{\rho}\eta_{xxxx} + gH\eta_{xx}~.
\end{equation}
A comparison between the quantum mechanical and the hydrodynamic frameworks may now be realized by the consistent conversion,
\begin{equation}
    \frac{\sigma H}{\rho} \rightarrow \frac{\hbar^2}{4m^2}, \quad gH \rightarrow \frac{V_0}{m}~,
\end{equation}
and summarized in table \eqref{tab:QM_GC}; the additional constant in the dispersion relation of the quantum system represents an additional potential in the wave equation, which does not exist in the shallow water wave equation.
\renewcommand{\arraystretch}{2}
\begin{table}
\centering
\caption{\label{tab:QM_GC}Comparison between the dispersion relation of RS and GC. We normalized the equations using the natural temporal and spatial frequencies, $\omega_n$ and $k_n$.}
\begin{tabular}{c|c|c}
\hline \hline
&Quantum Mechanics&Fluid Dynamics\\
\hline
DR&$\omega^2=\frac{\hbar^2}{4m^2}k^4+\frac{V_0}{m}k^2+\frac{V_0^2}{\hbar^2}$ & $\omega_F^2=\frac{\sigma H}{\rho} k_F^4+gHk_F^2$\\
$\omega_n$&$2\frac{V_0}{\hbar}$&$\sqrt{\frac{g^2H}{\sigma/\rho}}$\\
$k_n$&$2\frac{\sqrt{mV_0}}{\hbar}$ &$\sqrt{\frac{g}{\sigma/\rho}}$\\
norm-$DR$ & $\omega^2=k^4+k^2+1/4$ & $\omega^2=k^4+k^2$\\
norm-WE & 
$\xi_{tt} = -\xi_{xxxx} + \xi_{xx} - \frac{1}{4}\xi$ &
$\eta_{tt} = -\eta_{xxxx} + \eta_{xx}$\\
\hline \hline
\end{tabular}
\end{table}
Nevertheless, as shown below, the two equations are comparable at the limit of small potentials.  
In the limit of $V_0 = 0$ where the particle dynamics are independent of the value of the potential itself, the dispersion relation of the QM wave equation becomes $\omega^2 = \frac{\hbar^2}{4m^2}k^4$. 
The equivalent of this limit is the purely shallow fluid capillary dispersion relation, excluding the influence of gravity,
\begin{equation}
    \omega^2 = \frac{\sigma H}{\rho}k^4,
\end{equation}
Hence, an analogy of the shallow fluid purely capillary waves and the Schr\"{o}dinger equation is found, where $\omega = \frac{\hbar}{2m}k^2$.
In this analogy, the constant density of water waves is equivalent to the particle rest mass, and the surface tension $\sigma$ plays the role of the Planck constant $\hbar$.
Thus, by excluding gravity from the hydrodynamic system, we may obtain an exact analog to the dispersion relation of a free particle in quantum mechanics. 

We proceed by investigating the influence of constant potentials on particle statistics.
Henceforth, we shall use the RS equation in the following theoretical model.
Similar to \cite{dagan2020hydrodynamic, durey2020hydrodynamic}, the particle is modeled as a localized time-periodic disturbance of the wave field. 
The general forced RS equation is
\begin{equation}\label{wave equation plus force}
        \xi_{tt}=-\frac{\hbar^2}{4m^2}\xi_{xxxx}+\frac{V_0}{m}\xi_{xx}-\frac{V_0^2}{\hbar^2}\xi+F(x,x_p,t)~,
\end{equation}
where $F$ is a forcing term comprised of a spatial function localized at particle position $x_p$, multiplied by a time-dependent periodic function,   
\begin{equation}
    F_{\omega_n,k_n}(x,x_p,t) = - \gamma f(t) g\left(x-x_p(t)\right)~.
\end{equation}
Here, $f_{\omega_n}(t)=\sin{(2\omega_n t)}$ and 
\[ g_{k_n}(x-x_p)=\frac{1}{a\sqrt{\pi}}\exp\bigg[-\left(\frac{x-x_p}{a} \right)^2\bigg ] \] 
is a normalized Gaussian function.
We shall assume that particles respond to resonant interactions at the Compton frequency~\cite{bush2020hydrodynamic}, which is manifested through the potential of the RS wave equation. 
Hence, changing the potential would also change the natural frequencies of the system, $\omega_n, k_n$, correspondingly.
The forcing frequency is then set to twice the natural frequency, following~\cite{dagan2020hydrodynamic, durey2020hydrodynamic}. 
The characteristic width of the Gaussian represents the particle effect on the scale of the wave's natural wavelength, where $a=\frac{1}{2}\lambda_n=\pi/k_n$. $\gamma$ is a constant which, due to scaling analysis, takes the form of $\gamma_{\omega_n,k_n}=\gamma_0 \frac{\omega_n^2}{k_n^2}$, where $\gamma_0$ is dimensionless.

To examine the behavior of the particle under different excitation energies, we may change the potential of the system $V'$ relative to $V_0$, such that $V'=\epsilon V_0$ and the characteristic frequencies $\omega'=\epsilon\omega_n$ and $k'=\sqrt{\epsilon}k_n$ will change accordingly.  
To achieve resonance interaction between the particle and wave at its natural frequency (similar to \cite{dagan2020hydrodynamic, durey2020hydrodynamic}), the current model changes both the forcing energy and the potential of the system with the appropriate natural frequency and wave number.
As a result of the periodic disturbance in a bounded domain, this model continuously exerts energy into the system; in the hydrodynamic system, waves are damped due to viscosity effects, and higher amplitudes occur in the vicinity of the particle. To expel energy out of the system, we add the linear damping term, $b u_t$.
The influence of this coefficient on the system is kept small by minimizing the coefficient $b$. 
Using the normalization factors,  $x=\tilde{x}/k_n$, $t=\tilde{t}/\omega_n$ and $\xi=\tilde{\xi}/k_n$, the full dimensionless wave equation under a modified potential $V'$ may be written as
\begin{equation}
        \xi_{tt} + b\xi_t = -\xi_{xxxx}+\epsilon\xi_{xx} - \frac{\epsilon^2}{4}\xi + F_{\omega',k'}(x,x_p,t),
\end{equation}
where ``tilde" is omitted for clarity.
Note that the potential and particle effects vanish at the limit $\epsilon=0$. Analogously, without a particle, the fluid bath in HQA remains flat when below the Faraday threshold. Here, the initial conditions of the field are $\xi=0$, $\xi_t=0$ everywhere, so with the absence of a particle effect, the field remains trivial, $\xi=0$. We determine a reference energy to the problem, $\epsilon=0$. If we change the potential relative to this reference, we expect a change in the particle's kinetic energy.
Markedly, in the limit $\epsilon \ll 1$, we may neglect quadrature terms, and the one-dimensional shallow fluid GC wave equation and the RS equation are identical.

Until now, we only described the wave field and modeled its response to an oscillating particle; the effect of the waves on the particle motion remains to be explained. 
The equation of motion describing the particle's horizontal displacement $x_p$ may be modeled as in \cite{oza2013trajectory},
\begin{equation}
    m\ddot x_p = -D\dot x_p - 
    f\nabla \eta(x_p,t)~.
\end{equation}
In this model, a particle is driven by wave gradients proportional to the local wave field $\eta(x,t)$, resisted by a linear drag, with a constant drag coefficient $D$.
$f$ is a constant wave coupling parameter.
Our previous studies of classical dynamics have shown how particles may exhibit anomalous diffusion~\cite{wang2024brownian} and particle clustering under the influence of shear forces, periodic vortex flows~\cite{dagan2017similarity, dagan2017particle, avni2023droplet1, avni2023droplet2} and oscillatory flows~\cite{dagan2021settling}. However, unlike the Langevin equation in which a random forcing term may lead to anomalous diffusion, the present model assumes deterministic wave gradients, from which complex nonlinear dynamics and particle diffusion and clustering may be realized. 
We further assume that particle inertia may be neglected~\cite{dagan2020hydrodynamic}, and the normalized guiding equation takes the dimensionless form,
\begin{equation} \label{dynamic}
    \Dot{x}_p = -\alpha \frac{\partial \xi}{\partial x}\Big|_{x_p}~.
\end{equation}
Hence, the evolving wave field continuously guides the particle, and its velocity is directly proportional to the wave gradients.
This method allows us to isolate the effect of wave gradients on the evolution of both random and ordered trajectories. 
\begin{figure}
    \centering
    \includegraphics[width=0.49\textwidth]{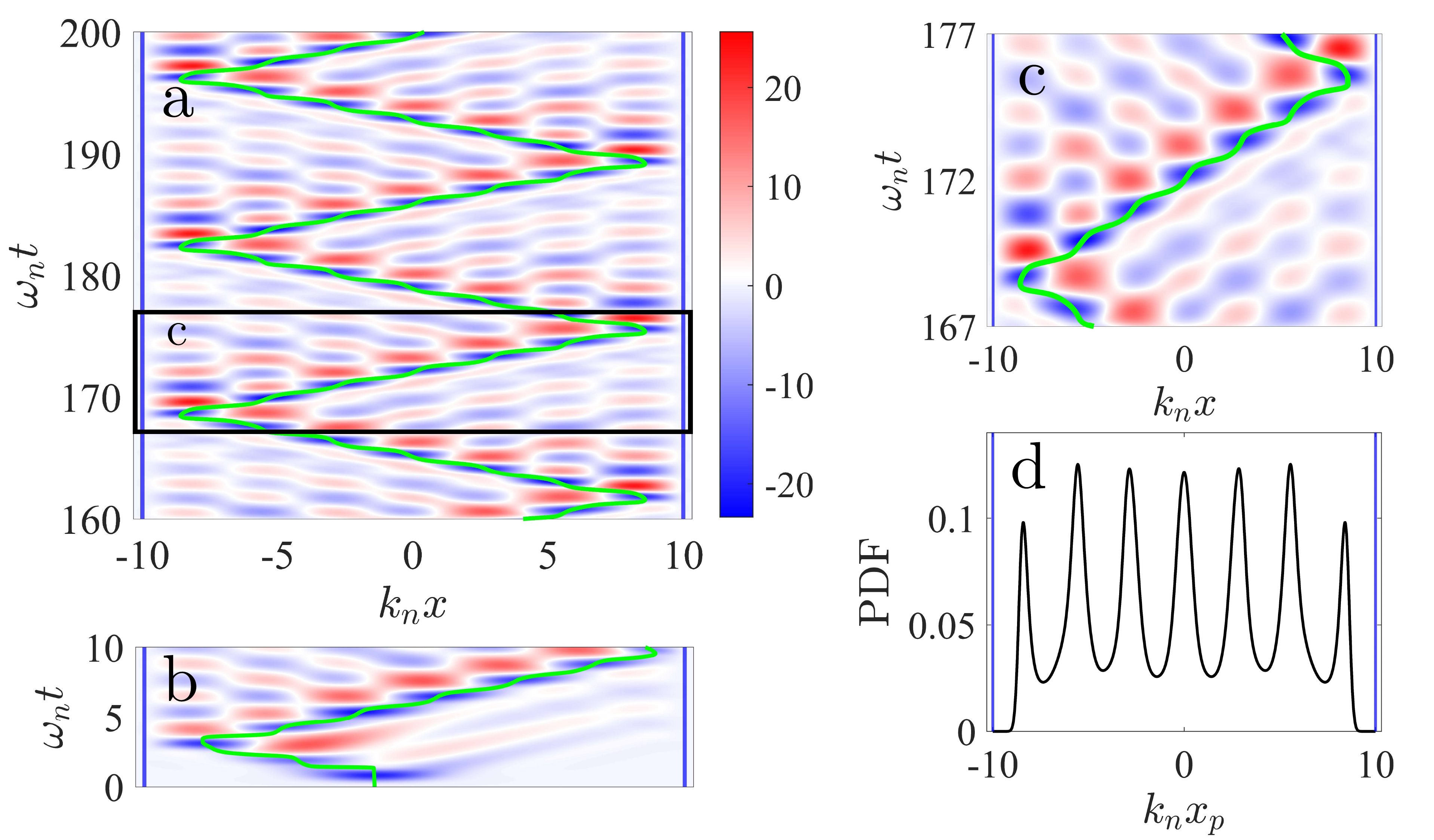}
    \caption{\label{fig:trajectory_and_pdf} (a) Spatio-temporal map section of the particle trajectory and its wavefield, for normalized potential $\epsilon=2.73$. (b) particle initial movement (c) Zoom-in of (a) one particle passing from side to side. (d) Projection of the trajectory on the particle's position PDF. The five main peaks determine the mode associated with the particle's kinetic energy.}
\end{figure}
Here, $\alpha$ is a free parameter, representing the extent to which the waves affect the particles. 
In our coupled particle-wave model, we initially place a stationary particle at an arbitrary location in a potential well, which generates waves centered about its initial location. 
Excitation of motion is observed when waves reflected from the potential boundaries break the wave symmetry about the particle location.

Notably, for different values of $\epsilon$, the waveform and the corresponding particle motion exhibit different characteristics, generally divided into two distinct modes. 
The first results in a periodic coherent particle trajectory, as shown in Fig.~\ref{fig:trajectory_and_pdf}. 
Here, a particle placed at an arbitrary location is set into motion due to local asymmetric field gradients.
In this case, the spatial probability density function of the particle location reveals five distinct peaks, suggesting quantum-like statistics.
The particle is locked into a quasi-steady motion, exhibiting inline oscillations similar to the motion reported by~\citet{dagan2020hydrodynamic}, then flips direction at the potential walls due to the spatial extent of its guiding wave. 
The second mode is characterized by irregular non-periodic motion, for which ordered inline motion is not apparent, and the emergent statistics do not have a noticeable coherent structure. 
All the spatially periodic modes of the real Schr\"{o}dinger equation are the modes of the complex Schr\"{o}dinger equation of free particles in a box,
\begin{equation}
    k_n = \sqrt{\frac{2m(E_n-V_0)}{\hbar^2}} = \pi n/L~,
\end{equation}
whereas other additional modes correspond to spatially diverging and decaying exponents.

To further explore the shallow water analogy, the probability density function (PDF) of the particle's location extracted from relatively long simulations (integrated over $t=30,000/\omega_n$) is calculated.
Figure \ref{fig:trajectory_and_pdf} presents the PDF of multiple long simulations. Each simulation was run for a specific normalized potential $\epsilon$, where the motion of the particle changes according to the varying potentials, which is reflected in the particle PDF. 
Figure~\ref{fig:trajectory_and_pdf} (d) shows that the particle location distribution in the box is described by five main and two smaller peaks. The smaller peaks represent the effective borders of the particle motion, where the particle changes its direction. Note that the flip in direction does not occur at the potential barrier but slightly inside the potential well, decreasing its effective length. 
The particle statistical distribution between the effective boundaries describes the particle's kinetic energy mode. 

By increasing the system's potential, relative to $\epsilon=0$, the particle's kinetic energy is increasing, and the PDF of finding the particle reveals an increasing number of peaks - interpreted here as hydrodynamic spatial modes.

We shall now use the present hydrodynamic analogy to study the influence of the potential change $\epsilon$ on the particle kinetic energy and the deterministic transition between modes, for which there is no parallel mechanism in quantum mechanics. Here, instead of changing the particle's energy, we realize a continuous change of the potential. In shallow water waves, this could be conducted by simply changing the water depth.

\begin{figure}
    \centering
    \includegraphics[width=0.49\textwidth]{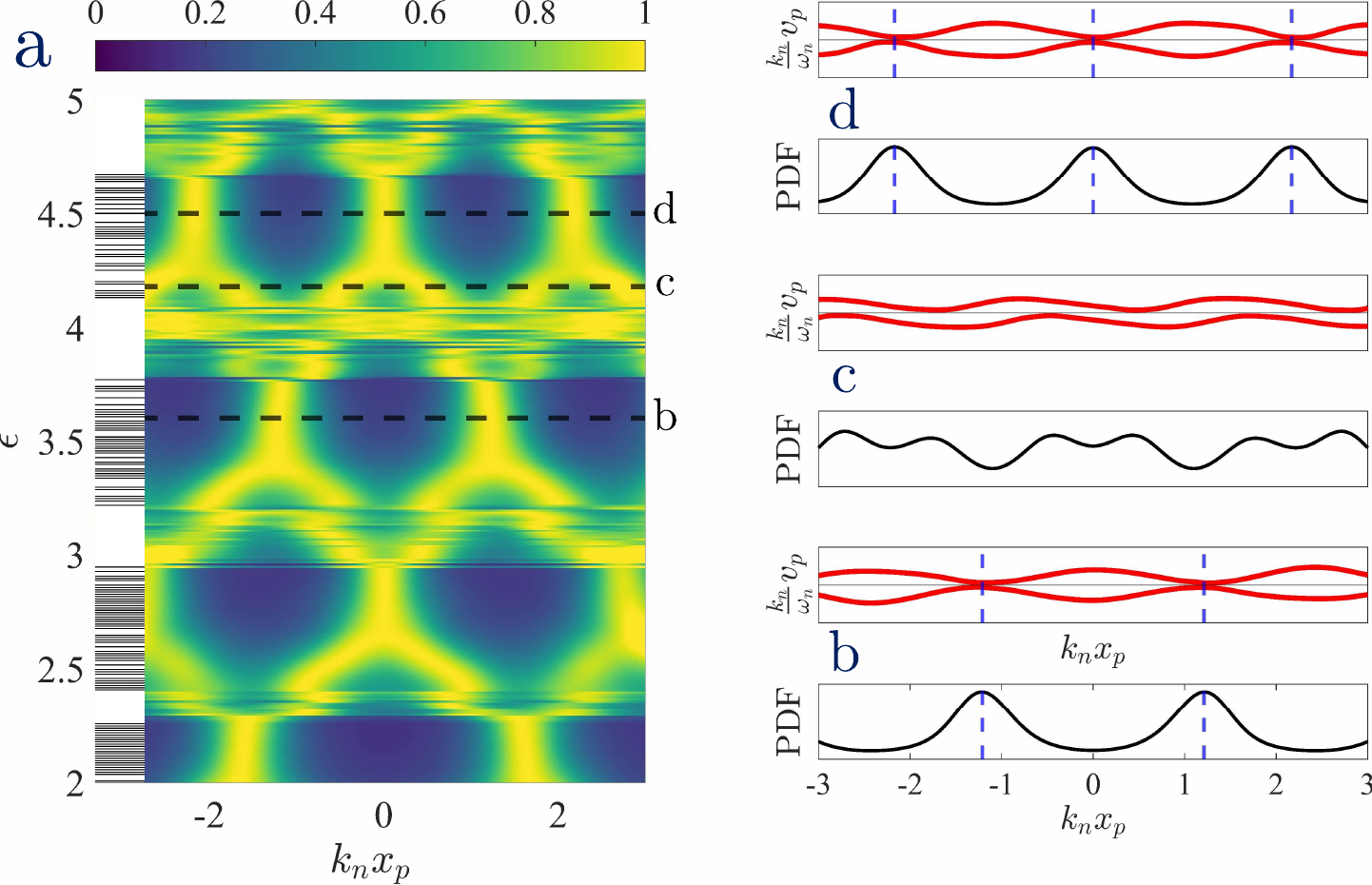}
    \caption{\label{fig:PDF_and_zoom} (a) The normalized PDF of the particle's location $\frac{PDF(\epsilon)}{PDF_{max}(\epsilon)}$ as a function of $\epsilon$, the black horizontal lines indicate that the particle stayed in the box throughout all simulation. The black lines of (b),(c), and (d) are also the PDF for specific $\epsilon$ values marked by dashed lines in (a). For each $\epsilon$ of (b),(c) and (d) the periodic phase space of $\left(k_nx_p(t),\frac{k_n}{\omega_n}v_p(t)\right)$ is plotted in red.}
\end{figure}

After an initial transient, the particle repeats its trajectory, and a periodic picture of the trajectory is revealed in the phase space in Fig.~\ref{fig:PDF_and_zoom}. 
Figure~\ref{fig:PDF_and_zoom}(a) shows the normalized PDF of the particle's location. The black lines on the left indicate whether the particle continued its periodic motion inside the box throughout the simulation. 
These regions are associated with coherent modes with clear peaks (color-coded by their PDF values), representing the modes of the particle in the box at a particular energy. 
On the other hand, the white areas represent transitions between the clear peaks. 
A bifurcation is observed during the transition, where each distinct peak is split into two. 
Further increasing the energy, these split peaks merge again and form a new stable mode, doubling the spatial periodicity of previous peaks.
This recurring process is observed, giving rise to the appearance of discrete new peaks as the energy increases.
The regions where the particle tends to stay stable in the box performing periodic movement are synchronized with the potential energy $\epsilon$ and the size of the box $L$.
This can be seen in Fig.\ref{fig:PDF_and_zoom} b-d, where the PDFs of three clear transitional states are plotted. For each one of the three PDFs, the periodic phase space is shown. In Fig.~\ref{fig:PDF_and_zoom}c, we can see a periodic motion with a period of two crossings for a particle traveling between the right wall back to the right wall again, compared to Fig.\ref{fig:PDF_and_zoom}b and d, where only one crossing is observed. Thus, the PDF of Fig.\ref{fig:PDF_and_zoom}c presents the transition and re-connection between the two branches. In Fig.\ref{fig:PDF_and_zoom}b,d, the particle slows down roughly at the same locations throughout the periodic motion, and as a result, a coherent eigenmode is formed. As we will see in the next section, the coherent modes of the hydrodynamically-inspired model are the modes of the known particle in a box system.

\begin{figure}
    \centering
    \includegraphics[width=0.4\textwidth]{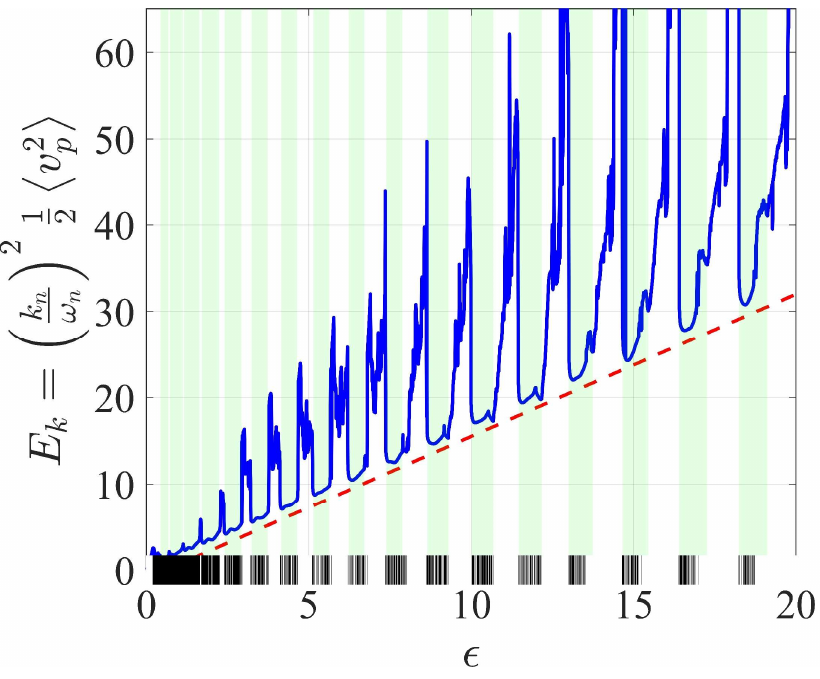}
    \caption{\label{fig:P_v_and_E_n1} The mean normalized kinetic energy $E_k$ as a function of the normalized potential $\epsilon$. The green vertical areas are the discrete $\epsilon_n$ corresponding to the kinetic energy's local stable regions. The black vertical lines indicate that the particle stayed in the box throughout the simulation.}
\end{figure}

\begin{figure}
    \centering
    \includegraphics[width=0.38\textwidth]{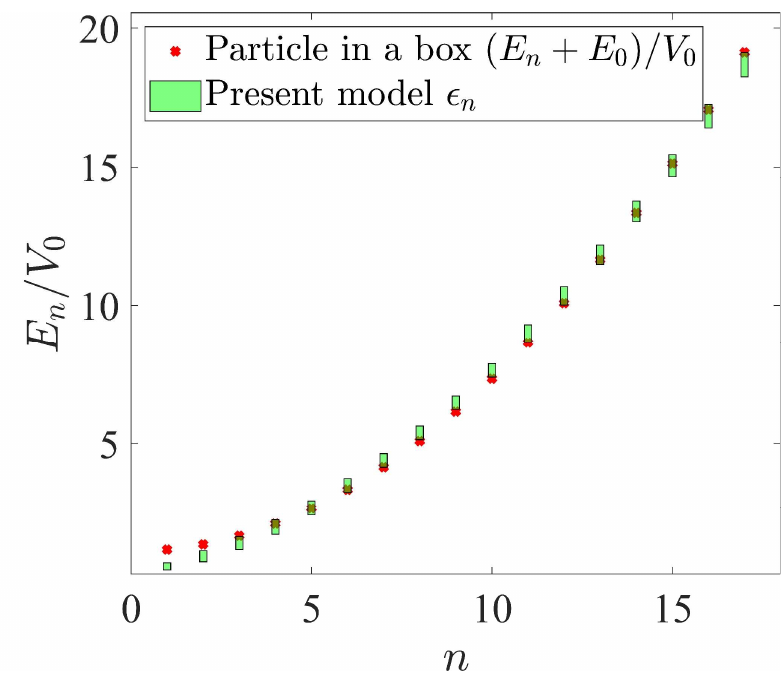}\caption{\label{fig:P_v_and_E_n2}Comparison between the normalized particle in a box energies and the present model energies $\epsilon_n$ chosen according to (a).}
\end{figure}

In contrast to the standard interpretation of quantum mechanics, in the present model, the particle's momentum is known at each spatiotemporal point along its trajectory. Therefore, we can extract its instantaneous and mean kinetic energy. The mean momentum $\braket{p}$ for each simulation with a different value of $\epsilon$ is zero due to the symmetry of the problem with respect to $x=0$. Still, we find it instructive to use the second moment $\braket{p^2}$ of each simulation. In the QM system, if the quantum state $\ket{\psi_n}$ is an eigenfunction of the Hamiltonian, the energy of the free particle is:
\begin{align} \label{E_n_P}
    E_n =\Braket{\psi_n|\frac{p^2}{2m}|\psi_n}= \frac{\hbar^2\pi^2}{2mL^2}n^2 
    = V_0 \frac{2\pi^2}{\tilde{L}^2}n^2~,
\end{align}
where $E_n$ is the corresponding eigenenergy, and $\tilde{L}=k_nL$ is the normalised box size. Here, in contrast to the standard QM interpretation, we continuously change the normalized potential $\epsilon$ through the different modes. Figure \ref{fig:P_v_and_E_n1} presents the mean kinetic energy of the hydrodynamic particle  
\[ E_k=\left(\frac{k_n}{\omega_nm} \right)^2\frac{\Braket{p^2}}{2}=\left(\frac{k_n}{\omega_n} \right)^2\frac{\Braket{v_p^2}}{2} \] 
as a function of $\epsilon$.
The mean kinetic energy increases with an increase in the potential while fluctuating in a rather chaotic manner. 
However, we find that the minima of the mean kinetic energy are proportional to the potential parameter $\epsilon$, with a chaotic-like region between every smooth minimum region. 
These stable regions corresponding to the linear growth are denoted in green stripes in Fig.\ref{fig:P_v_and_E_n1}.
In HQA, the statistical signature is obtained due to the spontaneous, chaotic transitions between unstable orbits, which are the peaks in the emergent droplet's PDF \cite{harris2014droplets,harris2013wavelike, cristea2018walking}. In Fig.\ref{fig:P_v_and_E_n1}, each stable region corresponds to a discrete energy level.

We may conclude that the emerging statistics of this classical framework stabilize according to the particle eigenmodes. The energy $\epsilon_n$ that corresponds to each mode is compared with the eigenenergy of the particle in a box system \eqref{E_n_P}, and presented here in Fig.\ref{fig:P_v_and_E_n2}, where we use the effective box size of $L=L_{eff}=17.79/k_n$.
Markedly, we observe an excellent match between the discrete energy set of the quantum system and the stable energies selected from a continuous domain of energies in the hydrodynamic model.
\begin{figure}
    \centering
    \includegraphics[width=0.49\textwidth]{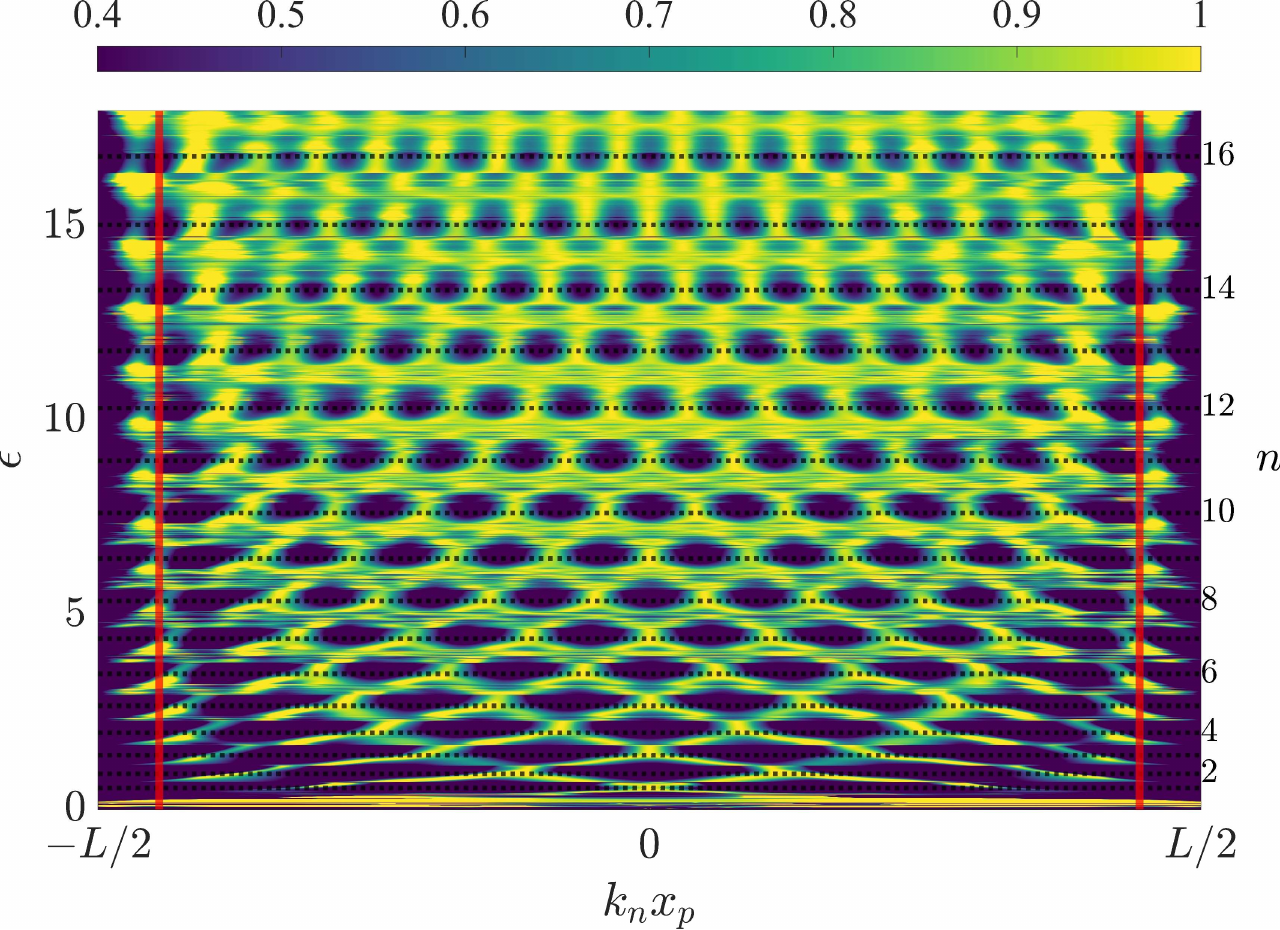}
    \caption{\label{fig:PDF_and_phi_mean} The normalized location PDF, $\frac{PDF(\epsilon)}{PDF_{max}(\epsilon)}$, in the box as a function of the normalized potential $\epsilon$. The black horizontal dashed lines are the energy levels $\epsilon_n$, where $n=1,..,16$, and the red vertical lines are the boundaries of the effective box.}
\end{figure}

The main result of the present study is summarized in Fig.~\ref{fig:PDF_and_phi_mean}. A map of the normalized PDF of the particle's location extracted from multiple simulations of different potentials $\epsilon$ reveals the emergence of discrete quantum-like statistics of classical particles driven by shallow water waves. The black horizontal dotted lines denote the energy levels found by the stable continuous region of the mean kinetic energy, $\epsilon_n$, namely the kinetic energy minima of the PDF. 
The vertical red lines represent the box's effective (numerical) boundaries, used for calculating the expected eigenenergies by equation \eqref{E_n_P}.
With this model, we can determine the discrete set of energies of particles in potential well quantum systems as well as relate its energy to its eigenmode manifested through the location PDF.
We thus propose a new, fully classical, deterministic mechanism for continuous transition between eigenmodes in quantum mechanics that also reproduces the allowed quantum particle states.

\bibliography{apssamp, rspw}

\begin{thebibliography}{29}%
\makeatletter
\providecommand \@ifxundefined [1]{%
 \@ifx{#1\undefined}
}%
\providecommand \@ifnum [1]{%
 \ifnum #1\expandafter \@firstoftwo
 \else \expandafter \@secondoftwo
 \fi
}%
\providecommand \@ifx [1]{%
 \ifx #1\expandafter \@firstoftwo
 \else \expandafter \@secondoftwo
 \fi
}%
\providecommand \natexlab [1]{#1}%
\providecommand \enquote  [1]{``#1''}%
\providecommand \bibnamefont  [1]{#1}%
\providecommand \bibfnamefont [1]{#1}%
\providecommand \citenamefont [1]{#1}%
\providecommand \href@noop [0]{\@secondoftwo}%
\providecommand \href [0]{\begingroup \@sanitize@url \@href}%
\providecommand \@href[1]{\@@startlink{#1}\@@href}%
\providecommand \@@href[1]{\endgroup#1\@@endlink}%
\providecommand \@sanitize@url [0]{\catcode `\\12\catcode `\$12\catcode `\&12\catcode `\#12\catcode `\^12\catcode `\_12\catcode `\%12\relax}%
\providecommand \@@startlink[1]{}%
\providecommand \@@endlink[0]{}%
\providecommand \url  [0]{\begingroup\@sanitize@url \@url }%
\providecommand \@url [1]{\endgroup\@href {#1}{\urlprefix }}%
\providecommand \urlprefix  [0]{URL }%
\providecommand \Eprint [0]{\href }%
\providecommand \doibase [0]{https://doi.org/}%
\providecommand \selectlanguage [0]{\@gobble}%
\providecommand \bibinfo  [0]{\@secondoftwo}%
\providecommand \bibfield  [0]{\@secondoftwo}%
\providecommand \translation [1]{[#1]}%
\providecommand \BibitemOpen [0]{}%
\providecommand \bibitemStop [0]{}%
\providecommand \bibitemNoStop [0]{.\EOS\space}%
\providecommand \EOS [0]{\spacefactor3000\relax}%
\providecommand \BibitemShut  [1]{\csname bibitem#1\endcsname}%
\let\auto@bib@innerbib\@empty
\bibitem [{\citenamefont {Madelung}(1926)}]{Madelung1926}%
  \BibitemOpen
  \bibfield  {author} {\bibinfo {author} {\bibfnamefont {E.}~\bibnamefont {Madelung}},\ }\bibfield  {title} {\bibinfo {title} {Quantentheorie in hydrodynamischen form},\ }\href@noop {} {\bibfield  {journal} {\bibinfo  {journal} {Zts. F. Phys.}\ }\textbf {\bibinfo {volume} {40}},\ \bibinfo {pages} {322} (\bibinfo {year} {1926})}\BibitemShut {NoStop}%
\bibitem [{\citenamefont {Bohm}(1952)}]{Bohm1952}%
  \BibitemOpen
  \bibfield  {author} {\bibinfo {author} {\bibfnamefont {D.}~\bibnamefont {Bohm}},\ }\bibfield  {title} {\bibinfo {title} {A suggested interpretation of the quantum theory in terms of hidden variables},\ }\href@noop {} {\bibfield  {journal} {\bibinfo  {journal} {Phys. Rev.}\ }\textbf {\bibinfo {volume} {85}},\ \bibinfo {pages} {166} (\bibinfo {year} {1952})}\BibitemShut {NoStop}%
\bibitem [{\citenamefont {Couder}\ and\ \citenamefont {Fort}(2006)}]{Couder2006}%
  \BibitemOpen
  \bibfield  {author} {\bibinfo {author} {\bibfnamefont {Y.}~\bibnamefont {Couder}}\ and\ \bibinfo {author} {\bibfnamefont {E.}~\bibnamefont {Fort}},\ }\bibfield  {title} {\bibinfo {title} {Single particle diffraction and interference at a macroscopic scale},\ }\href@noop {} {\bibfield  {journal} {\bibinfo  {journal} {Phys. Rev. Lett.}\ }\textbf {\bibinfo {volume} {97}},\ \bibinfo {pages} {154101} (\bibinfo {year} {2006})}\BibitemShut {NoStop}%
\bibitem [{\citenamefont {Bush}\ and\ \citenamefont {Oza}(2020)}]{bush2020hydrodynamic}%
  \BibitemOpen
  \bibfield  {author} {\bibinfo {author} {\bibfnamefont {J.~W.}\ \bibnamefont {Bush}}\ and\ \bibinfo {author} {\bibfnamefont {A.~U.}\ \bibnamefont {Oza}},\ }\bibfield  {title} {\bibinfo {title} {Hydrodynamic quantum analogs},\ }\href@noop {} {\bibfield  {journal} {\bibinfo  {journal} {Reports on progress in physics}\ }\textbf {\bibinfo {volume} {84}},\ \bibinfo {pages} {017001} (\bibinfo {year} {2020})}\BibitemShut {NoStop}%
\bibitem [{\citenamefont {Eddi}\ \emph {et~al.}(2009)\citenamefont {Eddi}, \citenamefont {Fort}, \citenamefont {Moisy},\ and\ \citenamefont {Couder}}]{eddi2009unpredictable}%
  \BibitemOpen
  \bibfield  {author} {\bibinfo {author} {\bibfnamefont {A.}~\bibnamefont {Eddi}}, \bibinfo {author} {\bibfnamefont {E.}~\bibnamefont {Fort}}, \bibinfo {author} {\bibfnamefont {F.}~\bibnamefont {Moisy}},\ and\ \bibinfo {author} {\bibfnamefont {Y.}~\bibnamefont {Couder}},\ }\bibfield  {title} {\bibinfo {title} {Unpredictable tunneling of a classical wave-particle association},\ }\href@noop {} {\bibfield  {journal} {\bibinfo  {journal} {Physical review letters}\ }\textbf {\bibinfo {volume} {102}},\ \bibinfo {pages} {240401} (\bibinfo {year} {2009})}\BibitemShut {NoStop}%
\bibitem [{\citenamefont {Fort}\ \emph {et~al.}(2010)\citenamefont {Fort}, \citenamefont {Eddi}, \citenamefont {Boudaoud}, \citenamefont {Moukhtar},\ and\ \citenamefont {Couder}}]{fort2010path}%
  \BibitemOpen
  \bibfield  {author} {\bibinfo {author} {\bibfnamefont {E.}~\bibnamefont {Fort}}, \bibinfo {author} {\bibfnamefont {A.}~\bibnamefont {Eddi}}, \bibinfo {author} {\bibfnamefont {A.}~\bibnamefont {Boudaoud}}, \bibinfo {author} {\bibfnamefont {J.}~\bibnamefont {Moukhtar}},\ and\ \bibinfo {author} {\bibfnamefont {Y.}~\bibnamefont {Couder}},\ }\bibfield  {title} {\bibinfo {title} {Path-memory induced quantization of classical orbits},\ }\href@noop {} {\bibfield  {journal} {\bibinfo  {journal} {Proceedings of the National Academy of Sciences}\ }\textbf {\bibinfo {volume} {107}},\ \bibinfo {pages} {17515} (\bibinfo {year} {2010})}\BibitemShut {NoStop}%
\bibitem [{\citenamefont {Oza}\ \emph {et~al.}(2014)\citenamefont {Oza}, \citenamefont {Harris}, \citenamefont {Rosales},\ and\ \citenamefont {Bush}}]{oza2014pilot}%
  \BibitemOpen
  \bibfield  {author} {\bibinfo {author} {\bibfnamefont {A.~U.}\ \bibnamefont {Oza}}, \bibinfo {author} {\bibfnamefont {D.~M.}\ \bibnamefont {Harris}}, \bibinfo {author} {\bibfnamefont {R.~R.}\ \bibnamefont {Rosales}},\ and\ \bibinfo {author} {\bibfnamefont {J.~W.}\ \bibnamefont {Bush}},\ }\bibfield  {title} {\bibinfo {title} {Pilot-wave dynamics in a rotating frame: on the emergence of orbital quantization},\ }\href@noop {} {\bibfield  {journal} {\bibinfo  {journal} {Journal of fluid mechanics}\ }\textbf {\bibinfo {volume} {744}},\ \bibinfo {pages} {404} (\bibinfo {year} {2014})}\BibitemShut {NoStop}%
\bibitem [{\citenamefont {Perrard}\ \emph {et~al.}(2014)\citenamefont {Perrard}, \citenamefont {Labousse}, \citenamefont {Miskin}, \citenamefont {Fort},\ and\ \citenamefont {Couder}}]{perrard2014self}%
  \BibitemOpen
  \bibfield  {author} {\bibinfo {author} {\bibfnamefont {S.}~\bibnamefont {Perrard}}, \bibinfo {author} {\bibfnamefont {M.}~\bibnamefont {Labousse}}, \bibinfo {author} {\bibfnamefont {M.}~\bibnamefont {Miskin}}, \bibinfo {author} {\bibfnamefont {E.}~\bibnamefont {Fort}},\ and\ \bibinfo {author} {\bibfnamefont {Y.}~\bibnamefont {Couder}},\ }\bibfield  {title} {\bibinfo {title} {Self-organization into quantized eigenstates of a classical wave-driven particle},\ }\href@noop {} {\bibfield  {journal} {\bibinfo  {journal} {Nature communications}\ }\textbf {\bibinfo {volume} {5}},\ \bibinfo {pages} {1} (\bibinfo {year} {2014})}\BibitemShut {NoStop}%
\bibitem [{\citenamefont {Gilet}(2016)}]{gilet2016quantumlike}%
  \BibitemOpen
  \bibfield  {author} {\bibinfo {author} {\bibfnamefont {T.}~\bibnamefont {Gilet}},\ }\bibfield  {title} {\bibinfo {title} {Quantumlike statistics of deterministic wave-particle interactions in a circular cavity},\ }\href@noop {} {\bibfield  {journal} {\bibinfo  {journal} {Physical Review E}\ }\textbf {\bibinfo {volume} {93}},\ \bibinfo {pages} {042202} (\bibinfo {year} {2016})}\BibitemShut {NoStop}%
\bibitem [{\citenamefont {Cristea-Platon}\ \emph {et~al.}(2018)\citenamefont {Cristea-Platon}, \citenamefont {S{\'a}enz},\ and\ \citenamefont {Bush}}]{cristea2018walking}%
  \BibitemOpen
  \bibfield  {author} {\bibinfo {author} {\bibfnamefont {T.}~\bibnamefont {Cristea-Platon}}, \bibinfo {author} {\bibfnamefont {P.~J.}\ \bibnamefont {S{\'a}enz}},\ and\ \bibinfo {author} {\bibfnamefont {J.~W.}\ \bibnamefont {Bush}},\ }\bibfield  {title} {\bibinfo {title} {Walking droplets in a circular corral: Quantisation and chaos},\ }\href@noop {} {\bibfield  {journal} {\bibinfo  {journal} {Chaos: An Interdisciplinary Journal of Nonlinear Science}\ }\textbf {\bibinfo {volume} {28}},\ \bibinfo {pages} {096116} (\bibinfo {year} {2018})}\BibitemShut {NoStop}%
\bibitem [{\citenamefont {S{\'a}enz}\ \emph {et~al.}(2018)\citenamefont {S{\'a}enz}, \citenamefont {Cristea-Platon},\ and\ \citenamefont {Bush}}]{saenz2018statistical}%
  \BibitemOpen
  \bibfield  {author} {\bibinfo {author} {\bibfnamefont {P.~J.}\ \bibnamefont {S{\'a}enz}}, \bibinfo {author} {\bibfnamefont {T.}~\bibnamefont {Cristea-Platon}},\ and\ \bibinfo {author} {\bibfnamefont {J.~W.}\ \bibnamefont {Bush}},\ }\bibfield  {title} {\bibinfo {title} {Statistical projection effects in a hydrodynamic pilot-wave system},\ }\href@noop {} {\bibfield  {journal} {\bibinfo  {journal} {Nature Physics}\ }\textbf {\bibinfo {volume} {14}},\ \bibinfo {pages} {315} (\bibinfo {year} {2018})}\BibitemShut {NoStop}%
\bibitem [{\citenamefont {S{\'a}enz}\ \emph {et~al.}(2020)\citenamefont {S{\'a}enz}, \citenamefont {Cristea-Platon},\ and\ \citenamefont {Bush}}]{saenz2020hydrodynamic}%
  \BibitemOpen
  \bibfield  {author} {\bibinfo {author} {\bibfnamefont {P.~J.}\ \bibnamefont {S{\'a}enz}}, \bibinfo {author} {\bibfnamefont {T.}~\bibnamefont {Cristea-Platon}},\ and\ \bibinfo {author} {\bibfnamefont {J.~W.}\ \bibnamefont {Bush}},\ }\bibfield  {title} {\bibinfo {title} {A hydrodynamic analog of friedel oscillations},\ }\href@noop {} {\bibfield  {journal} {\bibinfo  {journal} {Science advances}\ }\textbf {\bibinfo {volume} {6}},\ \bibinfo {pages} {eaay9234} (\bibinfo {year} {2020})}\BibitemShut {NoStop}%
\bibitem [{\citenamefont {Shinbrot}(2019)}]{shinbrot2019dynamic}%
  \BibitemOpen
  \bibfield  {author} {\bibinfo {author} {\bibfnamefont {T.}~\bibnamefont {Shinbrot}},\ }\bibfield  {title} {\bibinfo {title} {Dynamic pilot wave bound states},\ }\href@noop {} {\bibfield  {journal} {\bibinfo  {journal} {Chaos: An Interdisciplinary Journal of Nonlinear Science}\ }\textbf {\bibinfo {volume} {29}},\ \bibinfo {pages} {113124} (\bibinfo {year} {2019})}\BibitemShut {NoStop}%
\bibitem [{\citenamefont {Borghesi}(2017)}]{Borghesi2017}%
  \BibitemOpen
  \bibfield  {author} {\bibinfo {author} {\bibfnamefont {C.}~\bibnamefont {Borghesi}},\ }\bibfield  {title} {\bibinfo {title} {Equivalent quantum equations in a system inspired by bouncing droplets experiments},\ }\href@noop {} {\bibfield  {journal} {\bibinfo  {journal} {Foundations of Physics}\ }\textbf {\bibinfo {volume} {47}},\ \bibinfo {pages} {933} (\bibinfo {year} {2017})}\BibitemShut {NoStop}%
\bibitem [{\citenamefont {Drezet}\ \emph {et~al.}(2020)\citenamefont {Drezet}, \citenamefont {Jamet}, \citenamefont {Bertschy}, \citenamefont {Ralko},\ and\ \citenamefont {Poulain}}]{Drezet2020}%
  \BibitemOpen
  \bibfield  {author} {\bibinfo {author} {\bibfnamefont {A.}~\bibnamefont {Drezet}}, \bibinfo {author} {\bibfnamefont {P.}~\bibnamefont {Jamet}}, \bibinfo {author} {\bibfnamefont {D.}~\bibnamefont {Bertschy}}, \bibinfo {author} {\bibfnamefont {A.}~\bibnamefont {Ralko}},\ and\ \bibinfo {author} {\bibfnamefont {C.}~\bibnamefont {Poulain}},\ }\bibfield  {title} {\bibinfo {title} {Mechanical analog of quantum bradyons and tachyons},\ }\href@noop {} {\bibfield  {journal} {\bibinfo  {journal} {Physical Review E}\ }\textbf {\bibinfo {volume} {102}},\ \bibinfo {pages} {1} (\bibinfo {year} {2020})}\BibitemShut {NoStop}%
\bibitem [{\citenamefont {Dagan}\ and\ \citenamefont {Bush}(2020)}]{dagan2020hydrodynamic}%
  \BibitemOpen
  \bibfield  {author} {\bibinfo {author} {\bibfnamefont {Y.}~\bibnamefont {Dagan}}\ and\ \bibinfo {author} {\bibfnamefont {J.~W.}\ \bibnamefont {Bush}},\ }\bibfield  {title} {\bibinfo {title} {Hydrodynamic quantum field theory: the free particle},\ }\href@noop {} {\bibfield  {journal} {\bibinfo  {journal} {Comptes Rendus. M{\'e}canique}\ }\textbf {\bibinfo {volume} {348}},\ \bibinfo {pages} {555} (\bibinfo {year} {2020})}\BibitemShut {NoStop}%
\bibitem [{\citenamefont {Durey}\ and\ \citenamefont {Bush}(2020)}]{durey2020hydrodynamic}%
  \BibitemOpen
  \bibfield  {author} {\bibinfo {author} {\bibfnamefont {M.}~\bibnamefont {Durey}}\ and\ \bibinfo {author} {\bibfnamefont {J.~W.}\ \bibnamefont {Bush}},\ }\bibfield  {title} {\bibinfo {title} {Hydrodynamic quantum field theory: The onset of particle motion and the form of the pilot wave},\ }\href@noop {} {\bibfield  {journal} {\bibinfo  {journal} {Frontiers in Physics}\ }\textbf {\bibinfo {volume} {8}},\ \bibinfo {pages} {300} (\bibinfo {year} {2020})}\BibitemShut {NoStop}%
\bibitem [{\citenamefont {Dagan}(2023)}]{dagan2023relativistic}%
  \BibitemOpen
  \bibfield  {author} {\bibinfo {author} {\bibfnamefont {Y.}~\bibnamefont {Dagan}},\ }\bibfield  {title} {\bibinfo {title} {Relativistic hydrodynamic interpretation of de broglie matter waves},\ }\href@noop {} {\bibfield  {journal} {\bibinfo  {journal} {Foundations of Physics}\ }\textbf {\bibinfo {volume} {53}},\ \bibinfo {pages} {1} (\bibinfo {year} {2023})}\BibitemShut {NoStop}%
\bibitem [{\citenamefont {Dagan}(2024)}]{2024pilotwave}%
  \BibitemOpen
  \bibfield  {author} {\bibinfo {author} {\bibfnamefont {Y.}~\bibnamefont {Dagan}},\ }\bibinfo {title} {Advances in pilot wave theory: From experiments to foundations}\ (\bibinfo  {publisher} {Springer Nature},\ \bibinfo {year} {2024})\ Chap.\ \bibinfo {chapter} {Hydrodynamically Inspired Pilot-Wave Theory: An Ensemble Interpretation}\BibitemShut {NoStop}%
\bibitem [{\citenamefont {Schr{\"o}dinger}(2003)}]{schrodinger2003collected}%
  \BibitemOpen
  \bibfield  {author} {\bibinfo {author} {\bibfnamefont {E.}~\bibnamefont {Schr{\"o}dinger}},\ }\href@noop {} {\emph {\bibinfo {title} {Collected papers on wave mechanics}}},\ Vol.\ \bibinfo {volume} {302}\ (\bibinfo  {publisher} {American Mathematical Soc.},\ \bibinfo {year} {2003})\BibitemShut {NoStop}%
\bibitem [{\citenamefont {Oza}\ \emph {et~al.}(2013)\citenamefont {Oza}, \citenamefont {Rosales},\ and\ \citenamefont {Bush}}]{oza2013trajectory}%
  \BibitemOpen
  \bibfield  {author} {\bibinfo {author} {\bibfnamefont {A.~U.}\ \bibnamefont {Oza}}, \bibinfo {author} {\bibfnamefont {R.~R.}\ \bibnamefont {Rosales}},\ and\ \bibinfo {author} {\bibfnamefont {J.~W.}\ \bibnamefont {Bush}},\ }\bibfield  {title} {\bibinfo {title} {A trajectory equation for walking droplets: hydrodynamic pilot-wave theory},\ }\href@noop {} {\bibfield  {journal} {\bibinfo  {journal} {Journal of Fluid Mechanics}\ }\textbf {\bibinfo {volume} {737}},\ \bibinfo {pages} {552} (\bibinfo {year} {2013})}\BibitemShut {NoStop}%
\bibitem [{\citenamefont {Wang}\ and\ \citenamefont {Dagan}(2024)}]{wang2024brownian}%
  \BibitemOpen
  \bibfield  {author} {\bibinfo {author} {\bibfnamefont {N.}~\bibnamefont {Wang}}\ and\ \bibinfo {author} {\bibfnamefont {Y.}~\bibnamefont {Dagan}},\ }\bibfield  {title} {\bibinfo {title} {Brownian particle diffusion in generalized polynomial shear flows},\ }\href@noop {} {\bibfield  {journal} {\bibinfo  {journal} {Physical Review E}\ }\textbf {\bibinfo {volume} {110}},\ \bibinfo {pages} {024117} (\bibinfo {year} {2024})}\BibitemShut {NoStop}%
\bibitem [{\citenamefont {Dagan}\ \emph {et~al.}(2017{\natexlab{a}})\citenamefont {Dagan}, \citenamefont {Greenberg},\ and\ \citenamefont {Katoshevski}}]{dagan2017similarity}%
  \BibitemOpen
  \bibfield  {author} {\bibinfo {author} {\bibfnamefont {Y.}~\bibnamefont {Dagan}}, \bibinfo {author} {\bibfnamefont {J.}~\bibnamefont {Greenberg}},\ and\ \bibinfo {author} {\bibfnamefont {D.}~\bibnamefont {Katoshevski}},\ }\bibfield  {title} {\bibinfo {title} {Similarity solutions for the evolution of polydisperse droplets in vortex flows},\ }\href@noop {} {\bibfield  {journal} {\bibinfo  {journal} {International Journal of Multiphase Flow}\ }\textbf {\bibinfo {volume} {97}},\ \bibinfo {pages} {1} (\bibinfo {year} {2017}{\natexlab{a}})}\BibitemShut {NoStop}%
\bibitem [{\citenamefont {Dagan}\ \emph {et~al.}(2017{\natexlab{b}})\citenamefont {Dagan}, \citenamefont {Katoshevski},\ and\ \citenamefont {Greenberg}}]{dagan2017particle}%
  \BibitemOpen
  \bibfield  {author} {\bibinfo {author} {\bibfnamefont {Y.}~\bibnamefont {Dagan}}, \bibinfo {author} {\bibfnamefont {D.}~\bibnamefont {Katoshevski}},\ and\ \bibinfo {author} {\bibfnamefont {J.~B.}\ \bibnamefont {Greenberg}},\ }\bibfield  {title} {\bibinfo {title} {Particle and droplet clustering in oscillatory vortical flows},\ }\href@noop {} {\bibfield  {journal} {\bibinfo  {journal} {Atomization and Sprays}\ }\textbf {\bibinfo {volume} {27}} (\bibinfo {year} {2017}{\natexlab{b}})}\BibitemShut {NoStop}%
\bibitem [{\citenamefont {Avni}\ and\ \citenamefont {Dagan}(2023{\natexlab{a}})}]{avni2023droplet1}%
  \BibitemOpen
  \bibfield  {author} {\bibinfo {author} {\bibfnamefont {O.}~\bibnamefont {Avni}}\ and\ \bibinfo {author} {\bibfnamefont {Y.}~\bibnamefont {Dagan}},\ }\bibfield  {title} {\bibinfo {title} {Droplet dynamics in burgers vortices. i. mass transport},\ }\href@noop {} {\bibfield  {journal} {\bibinfo  {journal} {Physical Review Fluids}\ }\textbf {\bibinfo {volume} {8}},\ \bibinfo {pages} {083604} (\bibinfo {year} {2023}{\natexlab{a}})}\BibitemShut {NoStop}%
\bibitem [{\citenamefont {Avni}\ and\ \citenamefont {Dagan}(2023{\natexlab{b}})}]{avni2023droplet2}%
  \BibitemOpen
  \bibfield  {author} {\bibinfo {author} {\bibfnamefont {O.}~\bibnamefont {Avni}}\ and\ \bibinfo {author} {\bibfnamefont {Y.}~\bibnamefont {Dagan}},\ }\bibfield  {title} {\bibinfo {title} {Droplet dynamics in burgers vortices. ii. heat transfer},\ }\href@noop {} {\bibfield  {journal} {\bibinfo  {journal} {Physical Review Fluids}\ }\textbf {\bibinfo {volume} {8}},\ \bibinfo {pages} {083605} (\bibinfo {year} {2023}{\natexlab{b}})}\BibitemShut {NoStop}%
\bibitem [{\citenamefont {Dagan}(2021)}]{dagan2021settling}%
  \BibitemOpen
  \bibfield  {author} {\bibinfo {author} {\bibfnamefont {Y.}~\bibnamefont {Dagan}},\ }\bibfield  {title} {\bibinfo {title} {Settling of particles in the vicinity of vortex flows},\ }\href@noop {} {\bibfield  {journal} {\bibinfo  {journal} {Atomization and Sprays}\ }\textbf {\bibinfo {volume} {31}} (\bibinfo {year} {2021})}\BibitemShut {NoStop}%
\bibitem [{\citenamefont {Harris}\ and\ \citenamefont {Bush}(2014)}]{harris2014droplets}%
  \BibitemOpen
  \bibfield  {author} {\bibinfo {author} {\bibfnamefont {D.~M.}\ \bibnamefont {Harris}}\ and\ \bibinfo {author} {\bibfnamefont {J.~W.}\ \bibnamefont {Bush}},\ }\bibfield  {title} {\bibinfo {title} {Droplets walking in a rotating frame: from quantized orbits to multimodal statistics},\ }\href@noop {} {\bibfield  {journal} {\bibinfo  {journal} {Journal of fluid mechanics}\ }\textbf {\bibinfo {volume} {739}},\ \bibinfo {pages} {444} (\bibinfo {year} {2014})}\BibitemShut {NoStop}%
\bibitem [{\citenamefont {Harris}\ \emph {et~al.}(2013)\citenamefont {Harris}, \citenamefont {Moukhtar}, \citenamefont {Fort}, \citenamefont {Couder},\ and\ \citenamefont {Bush}}]{harris2013wavelike}%
  \BibitemOpen
  \bibfield  {author} {\bibinfo {author} {\bibfnamefont {D.~M.}\ \bibnamefont {Harris}}, \bibinfo {author} {\bibfnamefont {J.}~\bibnamefont {Moukhtar}}, \bibinfo {author} {\bibfnamefont {E.}~\bibnamefont {Fort}}, \bibinfo {author} {\bibfnamefont {Y.}~\bibnamefont {Couder}},\ and\ \bibinfo {author} {\bibfnamefont {J.~W.}\ \bibnamefont {Bush}},\ }\bibfield  {title} {\bibinfo {title} {Wavelike statistics from pilot-wave dynamics in a circular corral},\ }\href@noop {} {\bibfield  {journal} {\bibinfo  {journal} {Physical Review E}\ }\textbf {\bibinfo {volume} {88}},\ \bibinfo {pages} {011001} (\bibinfo {year} {2013})}\BibitemShut {NoStop}%
\end{thebibliography}%

\end{document}